\documentclass[a4paper,11pt]{article}
\usepackage{pos}
\usepackage{subfigure}
\usepackage{slashed}
\usepackage{dsfont}
\usepackage{wrapfig}
\graphicspath{{Figures/}}

\newcommand{\Ml}{\mathcal{M}_{ud}}

\newcommand{\ml}{m_{ud}}

\newcommand{\Z}{\mathcal{Z}}

\newcommand{\Dp}{\slashed{D}(\mu_\text{I})}
\newcommand{\Dm}{\slashed{D}(-\mu_\text{I})}
\newcommand{\mui}{\mu_\text{I}}
\newcommand{\muB}{\mu_\text{B}}

\title{Searching for the BCS phase at nonzero isospin asymmetry}

\author[a]{Bastian B. Brandt}
\author*[b]{Francesca Cuteri}
\author[a]{Gergely Endr\H{o}di}

\affiliation[a]{Institute for Theoretical Physics, University of Bielefeld, D-33615 Bielefeld, Germany}

\affiliation[b]{Institute for Theoretical Physics, Goethe University, D-60438 Frankfurt am  Main, Germany}

\emailAdd{brandt@physik.uni-bielefeld.de}
\emailAdd{cuteri@itp.uni-frankfurt.de}
\emailAdd{endrodi@physik.uni-bielefeld.de}

\abstract{According to perturbation theory predictions, QCD matter in the zero-temperature, high-density limits of QCD at nonzero isospin chemical potential is expected to be in a superfluid Bardeen-Cooper-Schrieffer (BCS) phase of $u$ and $\bar{d}$ Cooper pairs.
It is also expected, on symmetry grounds, that such phase connects via an analytical crossover to the phase with Bose-Einstein condensation (BEC) of charged pions at $\mu_\text{I}\geq m_\pi/2$.
With lattice results, showing some indications that the deconfinement crossover also smoothly penetrates the BEC phase, the conjecture was made that the former connects continuously to the BEC-BCS crossover.
We compute the spectrum of the Dirac operator, and use generalized Banks-Casher relations, to test this conjecture and identify signatures of the superfluid BCS phase.}

\FullConference{%
 The 38th International Symposium on Lattice Field Theory, LATTICE2021
  26th-30th July, 2021
  Zoom/Gather@Massachusetts Institute of Technology
}

\begin{document}
\maketitle

\section{Introduction}

There exist physical systems/setups, such as non-central heavy-ion collisions, compact stars and their mergers, and likely the early Universe in some phase of its evolution, which consist of strongly interacting matter which is, borrowing condensed matter physics nomenclature, ``doped'' with an asymmetry in the number of up and down constituent quarks.
This asymmetry results in an excess of e.g.\ neutrons over protons or positively charged pions over negatively charged pions.
All mentioned systems made of isospin-asymmetric QCD matter are also characterized by some nonzero temperature, as well as by an even larger nonzero baryon density, i.e.\ they also carry an excess of matter over antimatter\footnote{We neglect electromagnetic fields in this study.
These are considered by another of our studies presented at this conference~\cite{ValoisLat2021}.}. 

To understand the thermodynamic properties of the mentioned systems, we would then need to map the phase diagram of QCD at least as a function of the temperature and baryon and isospin ``doping''.
Unfortunately, direct simulations of QCD in a setup characterized by a matter-antimatter asymmetry, encoded in a nonzero baryon chemical potential $\muB$, are hindered by the complex action problem.
Purely isospin asymmetric QCD matter, with a nonzero isospin density $n_\text{I}=n_\text{u}-n_\text{d}$ encoded in a nonzero isospin chemical potential $\mui=(\mu_\text{u}-\mu_\text{d})/2$, constitutes, instead, a setup which is amenable to direct Monte Carlo simulations.

It is certainly useful to study the QCD phase diagram in the $(T, \mui)$ plane at $\muB=0$, both as a first step towards a more complete description, and also to explain behaviour/properties of the mentioned physical systems, which result in particular from their isospin asymmetry (see e.g\ studies about the early Universe starting with large lepton flavour asymmetries~\cite{Wygas:2018otj,Middeldorf-Wygas:2020glx,Vovchenko:2020crk}).

\begin{figure}[t]
	\centering
   \subfigure[]{%
	\label{fig:phaseDiagr}\includegraphics[width=0.475\textwidth]{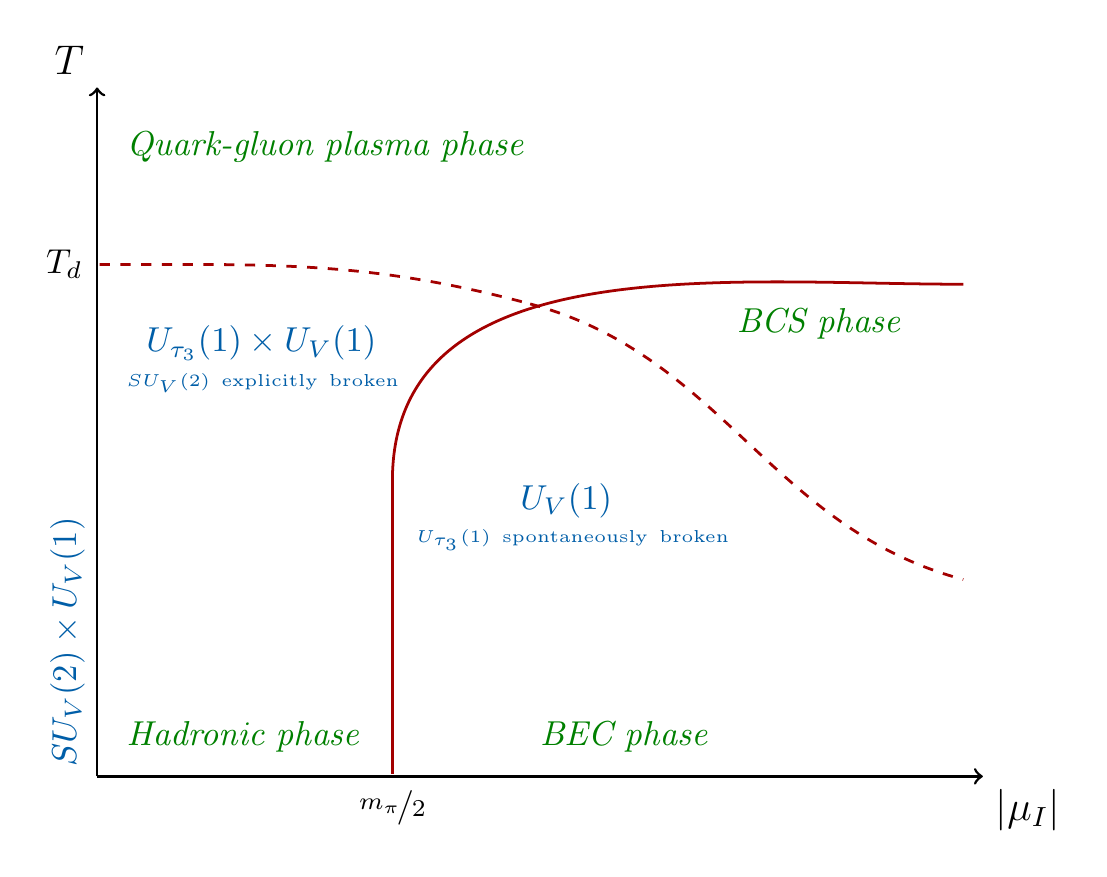}}
	\hfill
   \subfigure[]{%
	\label{fig:latticePhaseDiagr}\includegraphics[width=0.475\textwidth]{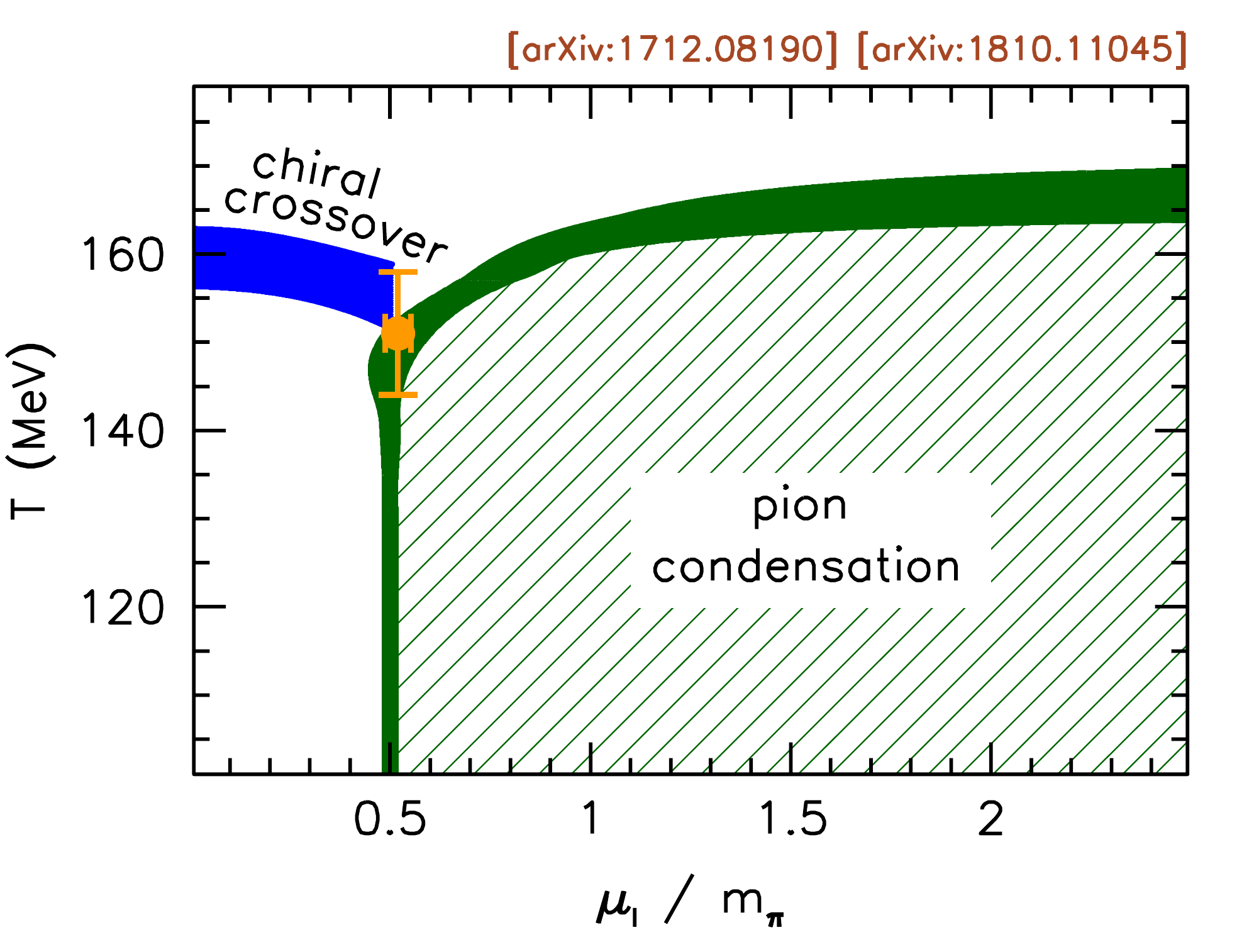}}
	\caption{Conjectured~\ref{fig:phaseDiagr}~\cite{Son:2000xc,Adhikari:2018cea} and measured~\ref{fig:latticePhaseDiagr}~\cite{Brandt:2017oyy,Brandt:2018omg} phase diagram of QCD at pure isospin chemical potential.}
\end{figure}

As anticipated by perturbation theory and model calculations~\cite{Son:2000xc,Adhikari:2018cea}, lattice simulations found~\cite{Brandt:2017oyy,Brandt:2018omg} (see Fig.~\ref{fig:latticePhaseDiagr}) for the QCD phase diagram in the $(T, \mui)$ plane, an interesting and complex structure with at least three phases.
The existence of a fourth BCS phase is also expected because perturbation theory, applicable in the limit $|\mui|\gg \Lambda_{QCD}$, predicts that the attractive gluon interaction forms pseudoscalar Cooper pairs of $u$ and $\bar d$ quarks at zero temperature~\cite{Son:2000xc}. 
Model calculations
also confirmed the existence of a BCS phase at nonzero temperature (see e.g.~\cite{Adhikari:2018cea}) and proposals for signatures of the BCS phase were made also in the context of two-color QCD~\cite{Boz:2019enj} and in chiral perturbation theory~\cite{Carignano:2016lxe}.
The transition between the BEC phase and the BCS phase is expected to be an analytic crossover, given that the symmetry breaking pattern is the same.
Moreover, as lattice simulations at nonzero isospin chemical potential have also revealed large values for the Polyakov loop within the BEC phase~\cite{Brandt:2017oyy}, considered as hints for a superconducting ground state with deconfined quarks, the conjectured phase diagram looks like Fig.~\ref{fig:phaseDiagr}.

In this proceedings we report on our progress in attempting to cross-check the existence of the BCS phase in Fig.~\ref{fig:phaseDiagr} and locate its boundaries by looking at the complex spectrum of the Dirac operator, in light of a Banks-Casher  type relation~\cite{Kanazawa:2013crb} (cf. Eq.~\eqref{eq:BC}).

\section{Simulation setup and observables}

Indicating by $S_{ud}=\bar\psi\Ml\,\psi$ the continuum action for the light quarks $\psi=(u,d)^\top$ in Euclidean spacetime, the fermion matrix $\Ml$ in the considered setup reads
\begin{equation}
\Ml = \gamma_\mu (\partial_\mu + i A_\mu)\, \mathds{1} + \ml \mathds{1} + \mui \gamma_4 \tau_3  + i \lambda \gamma_5 \tau_2\,,
\label{eq:Sud}
\end{equation}
with $A_\mu$ the gluon field, and $\tau_a$ the Pauli matrices.
The explicit symmetry breaking term $i \lambda \gamma_5 \tau_2$ with $\lambda$ referred to as pionic source coupling to the charged pion field is unphysical, but crucial in order to 
(1) enable the observation of the spontaneous breaking of the continuous $\mathrm{U}_{\tau_3}\!(1)$ symmetry and (2) regulate simulations in the BEC phase~\cite{Kogut:2002tm,Kogut:2002zg,Endrodi:2014lja}.
A safe $\lambda\to0$ extrapolation as devised in Ref.~\cite{Brandt:2017oyy} is necessary to obtain physical results.

For our measurements we consider 2+1-flavor QCD with $\mui>0$ and $\lambda>0$.
The quark masses are tuned to their physical values along the line of constant physics (LCP) from Ref.~\cite{Borsanyi:2010cj}, with the pion mass $m_\pi\approx 135\textmd{ MeV}$.
The Dirac operator is discretized employing the staggered formulation and the rooting procedure as in Refs.~\cite{Brandt:2017oyy,Brandt:2018omg} where the phase diagram shown in Fig~\ref{fig:latticePhaseDiagr} was mapped out.
The lattices considered so far are $N_s^3\times N_t$ lattices with $N_t\in\{6,12\}$ at various temperatures $T=1/(N_ta)$.

\begin{figure}[t]
	\centering
   \subfigure[]{%
	\label{fig:mui1}\includegraphics[width=0.325\textwidth]{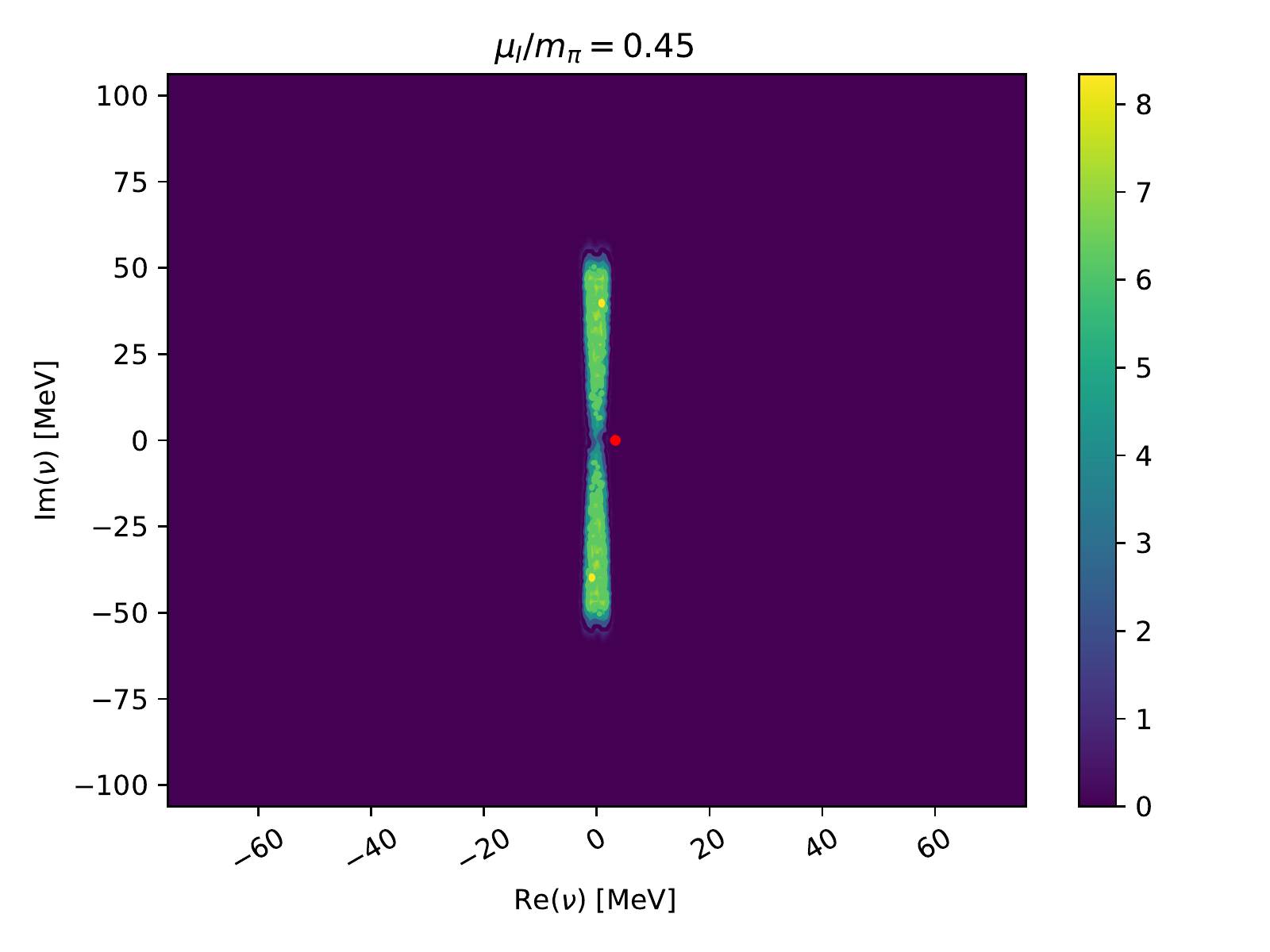}}
	\hfill
   \subfigure[]{%
	\label{fig:mui2}\includegraphics[width=0.325\textwidth]{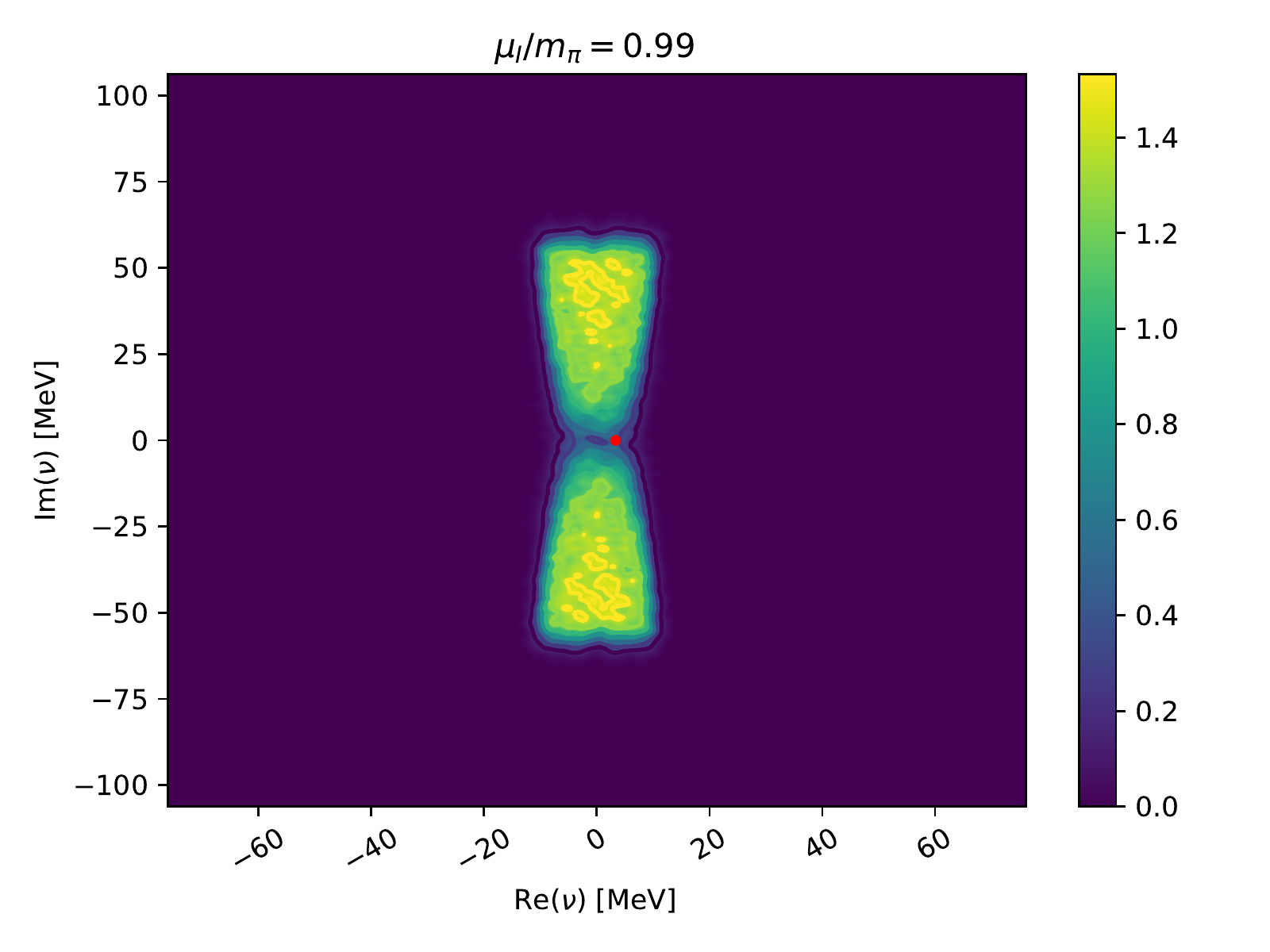}}
	\hfill
   \subfigure[]{%
	\label{fig:mui3}\includegraphics[width=0.325\textwidth]{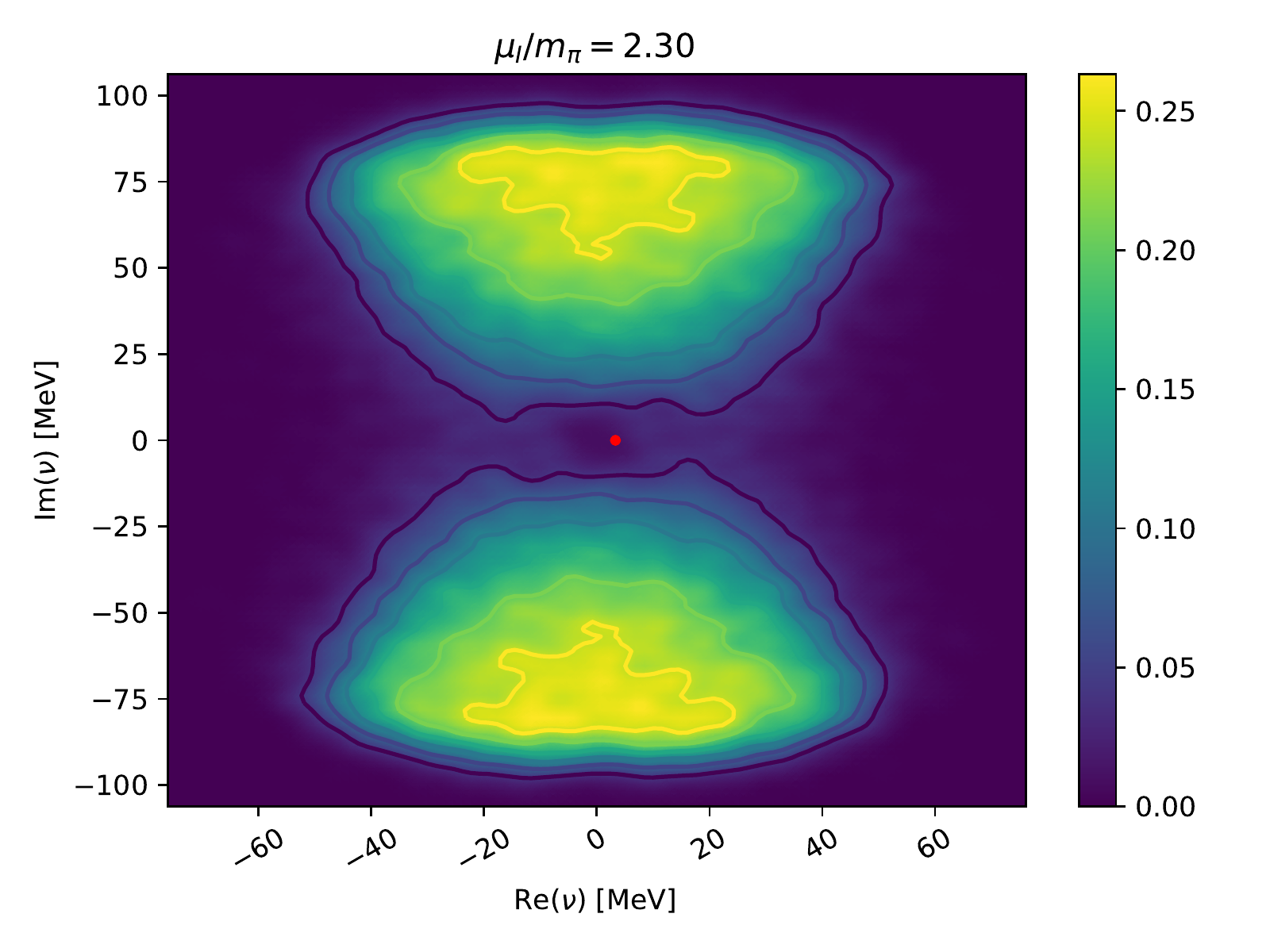}}
	\caption{\label{fig:mui}Contour plots of the complex spectrum of the Dirac operator as obtained for a $N_s=24$, $N_t=6$ lattice at $\lambda/\ml\sim0.29$ and $T=155$ MeV for isospin chemical potentials $\mui/m_\pi=0.45$~\ref{fig:mui1}, $\mui/m_\pi=0.99$~\ref{fig:mui2}, $\mui/m_\pi=2.30$~\ref{fig:mui3}.
	The red dot indicates $\ml + i \cdot 0$.}
\end{figure}

In the just described setup our observable of interest is the complex-eigenvalues spectrum of the massless Dirac operator $\Dp$.
For the up quark, the eigenproblem reads
\begin{equation}
\Dp \,\psi_n = \nu_n\,\psi_n\,,
\label{eq:Dpp}
\end{equation}
where the eigenvalues $\nu_n$ are complex numbers.
Using  chiral symmetry, i.e. $\Dp \eta_5 + \eta_5 \Dp = 0$ (with $\eta_5$ being the staggered equivalent of $\gamma_5$), and hermiticity, i.e. $\eta_5 \Dp\eta_5 = \Dm^\dagger$, the eigenproblem for the down quark can be obtained from Eq.~\eqref{eq:Dpp}, and it reads
\begin{equation}
{\widetilde\psi_n}^\dagger \,\Dm = {\widetilde\psi_n}^\dagger\,\nu_n^*\,,\quad\quad
\widetilde\psi_n = \eta_5\psi_n\,.
\label{eq:Dmm}
\end{equation}
Following Eqs.~(\ref{eq:Dpp}) and~(\ref{eq:Dmm}), for each eigenvalue in the up quark sector there is a complex conjugate pair in the down quark sector, see Fig.~\ref{fig:mui}.

The motivation behind our choice of observable lies in the existence of a Banks-Casher type relation
\begin{equation}
    \Delta^2 = \frac{2\pi^3}{9}\rho(0),
    \label{eq:BC}
\end{equation}
derived in Ref.~\cite{Kanazawa:2013crb} for the zero-temperature, high-isospin-density limits of QCD, which provides us with a prescription on how to obtain information on the BCS gap $\Delta$ from the density of the complex Dirac eigenvalues $\rho(\nu)$ evaluated, for the massless case, at the origin in the complex plane $\rho(0)$.

In Ref.~\cite{Kanazawa:2013crb}, the $T=0$ partition function $\Z(M)$ as a function of the quark mass matrix $M$ is considered both in the fundamental QCD-like theory and in the corresponding effective theory valid for $|\mui|\gg \Lambda_{QCD}$.
Taking suitable derivatives then yields in the fundamental theory an expression proportional to $\rho(0)$, and in the effective theory one proportional to $\Delta^2$ .
The Banks-Casher-type relation is obtained by identifying these results as in Eq.~\eqref{eq:BC}.

In our work we assume that a similar relation holds also in our setup with nonzero quark masses and temperatures.
To account for the nonzero quark masses in our simulations we evaluate the density $\rho(\nu)$ to $\ml+i \cdot 0$ rather than to zero neglecting, at first, possible corrections due to non-zero masses and temperatures.

\section{Results}

\begin{figure}[t]
	\centering
   \subfigure[]{%
	\label{fig:dataRhoMui}\includegraphics[width=0.4\textwidth]{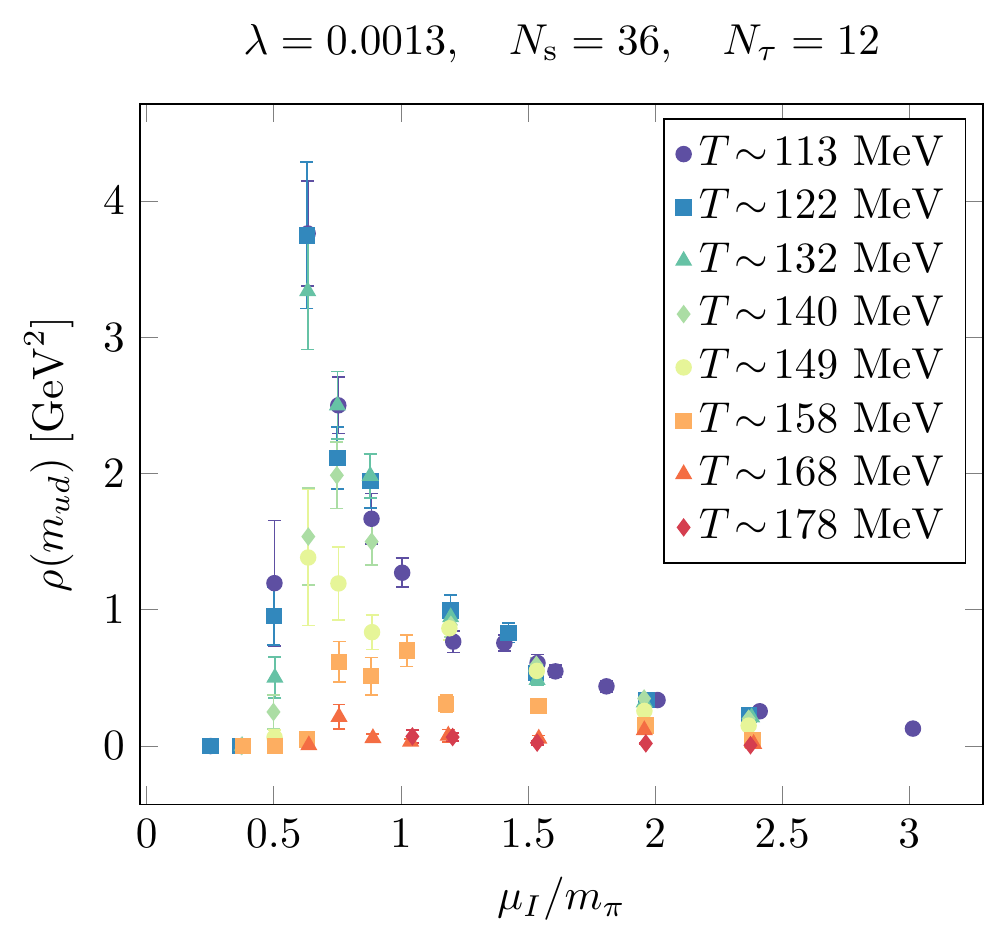}}
	\hfill
   \subfigure[]{%
	\label{fig:dataRhoMuiPhDiagr}\includegraphics[width=0.5\textwidth]{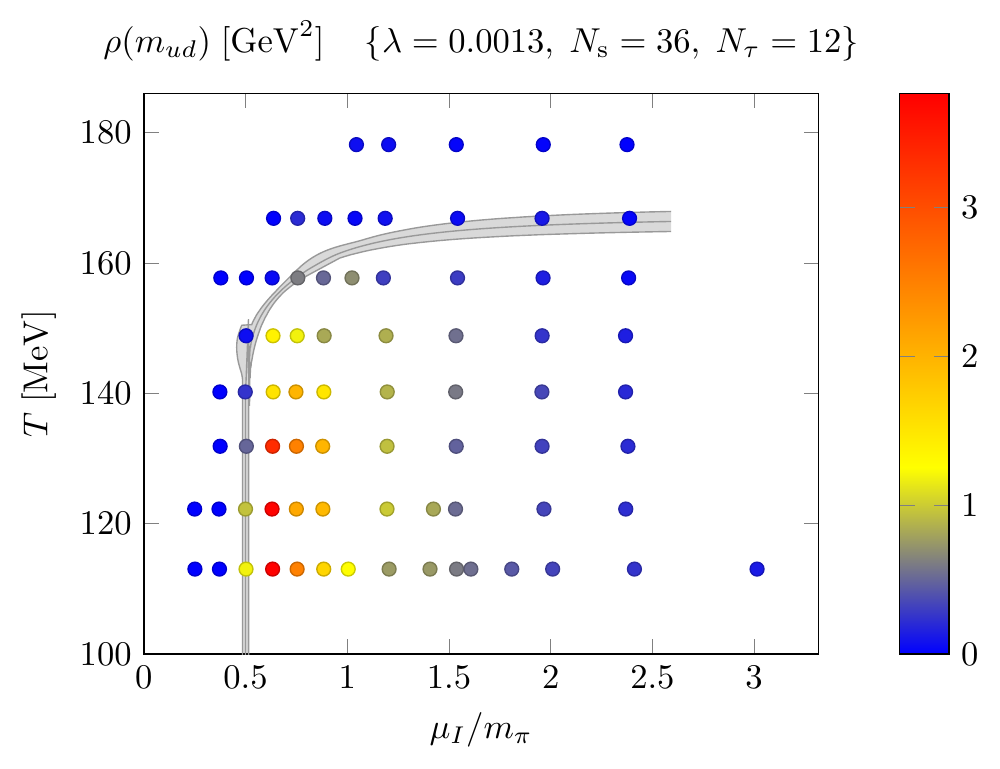}}
	\hfill
	\caption{\label{fig:density}\ref{fig:dataRhoMui} $\rho(\ml)$ plotted as a function of $\mui/m_\pi$ at eight different (approximate) values for $T$, for $N_t=12$ and the smallest considered value of the pionic source $\lambda=0.0013$.
	\ref{fig:dataRhoMuiPhDiagr} Density map plot for $\rho(\ml)$ in the $T$ {\it vs.} $\mui/m_\pi$ phase diagram for $N_t=12$, and $\lambda=0.0013$. The BEC boundary as obtained in Ref.~\cite{Brandt:2017oyy} for $N_t=12$ is also shown.}
\end{figure}

In order to solve the relevant eigenproblem in Eq.~(\ref{eq:Dpp}) we employ the Scalable Library for Eigenvalue Problem Computations (SLEPc)~\cite{slepc}, which is a software package for the solution of large sparse eigenproblems on parallel computers.
The solver used is a Krylov-Schur solver, whose implementation within SLEPc is suited for non-Hermitian problems.
We compute the closest to the origin (in modulo) $\sim150$ eigenvalues of the non-hermitian Dirac operator.

We evaluate $\rho(\ml)$ by using kernel density estimation (KDE), a non-parametric way to estimate the multivariate probability density function from the measured spectrum.
Such technique is implemented in the python library scikit-learn~\cite{scikit}, which we employ for the analysis.

By inspecting the density plots in Fig.~\ref{fig:mui}, it can be observed, how only for large enough $\mui$ values the complex spectrum gets wide enough in the real direction to encompass the red dot in Fig.~\ref{fig:mui} at  $\ml$, which results in $\rho(\ml)\neq0$.
At $\mui<m_\pi/2$ the eigenvalues are, instead, clustered along the imaginary axis and  $\rho(\ml)=0$.
At the largest simulated $\mui$ values, due to the drift of the eigenvalues away from the real axis, a decrease in $\rho(\ml)$ is observed.
A quantitative assessment of the impact of cutoff effects is necessary in order to tell to what extent they are responsible for the $\mui$-dependence of $\rho(\ml)$ at larger and larger $\mui$.

\begin{wrapfigure}{r}{7.2cm}
 \centering
 \includegraphics[width=7cm]{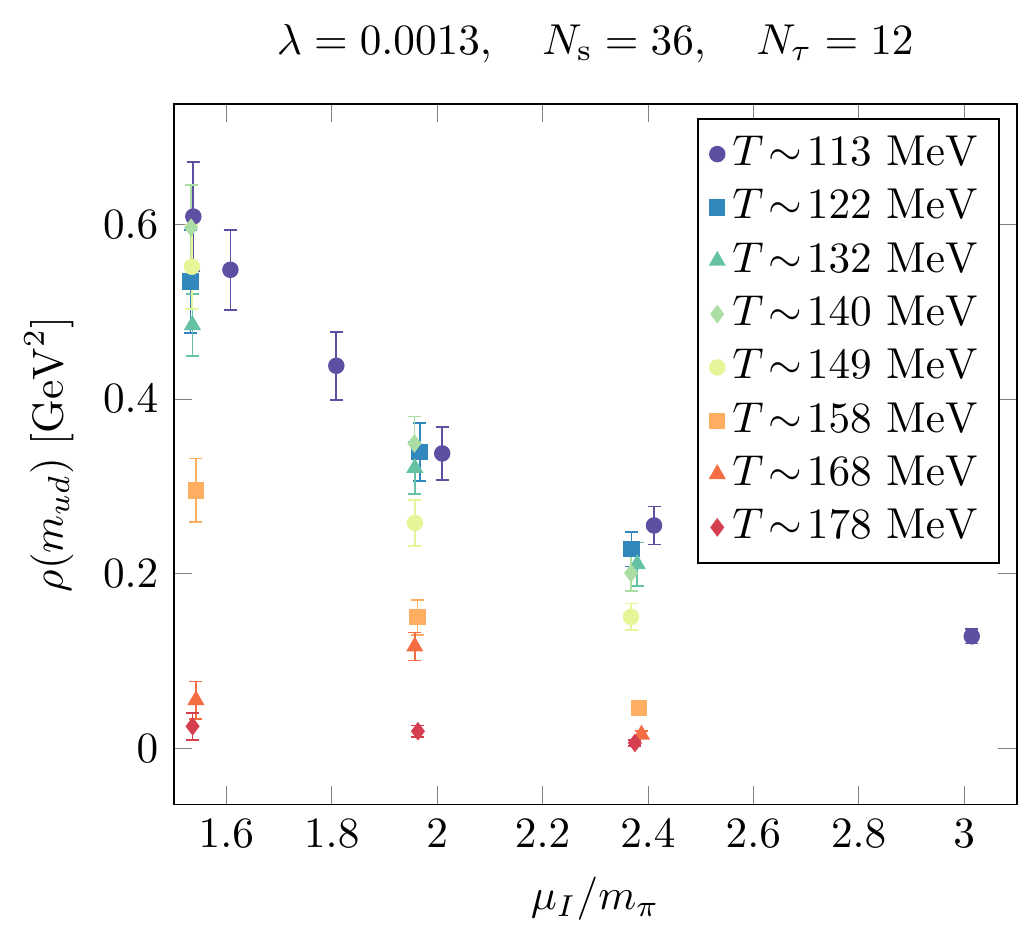}
 \caption{\label{fig:largeMui}Same as Fig.~\ref{fig:dataRhoMui}, but showing only the largest simulated $\mui$.}
\end{wrapfigure}

Quantitative results for the spectral density are shown in Fig.~\ref{fig:density}.
It is interesting to match the $\mui$- and $T$- dependence of $\rho(\ml)$ with the location of the boundary of the BEC phase as determined by the onset of the pion condensate $\Sigma_{\pi}$, see Fig.~\ref{fig:dataRhoMuiPhDiagr}.
What can be observed in Figs.~\ref{fig:dataRhoMui} and~\ref{fig:dataRhoMuiPhDiagr} is that the signal for the interpolated spectral density becomes nonzero exactly at $\mui^{\mathrm{BEC}}(T)$, that is at the location of the BEC phase boundary for the considered temperature.

Results also show the already mentioned drop in the values of the interpolated spectral density at larger values of $\mui$.
As already suggested lattice artefacts are expected to suppress $\rho(\ml)$, just as they do with $\Sigma_{\pi}$~\cite{Kogut:2002tm,Kogut:2002zg,Endrodi:2014lja}.
The results presented here have been obtained on the finest ensemble produced so far ($N_t=12$), for which $a\mui<0.25$ up until $\mui/m_\pi~2.5$, and therefore we expect lattice artefacts to be under control.
However, disentangling the signal for the BCS gap from discretization errors at even larger $\mui$ is difficult and requires a dedicated systematic study.

A closer inspection to the trend with $\mui$ of $\rho(\ml)$ at $\mui/m_\pi>1.5$ reveals (see Fig.~\ref{fig:largeMui}) that, statistically significant nonzero values are extracted also at the largest simulated $\mui$.

\begin{figure}[t]
	\centering
   \subfigure[]{%
	\label{fig:dataRhoMuiTDep}\includegraphics[width=0.4\textwidth]{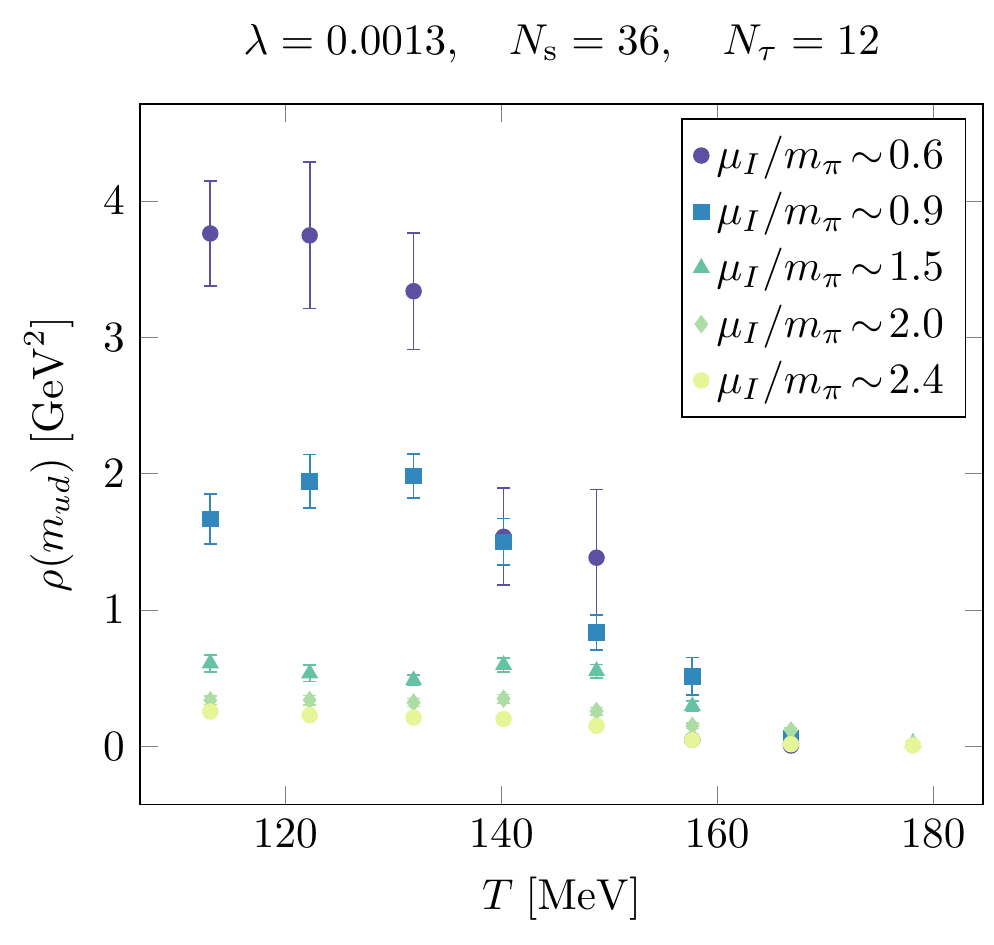}}
	\hspace{1cm}
   \subfigure[]{%
	\label{fig:dataRhoMuiLambdaDep}\includegraphics[width=0.413\textwidth]{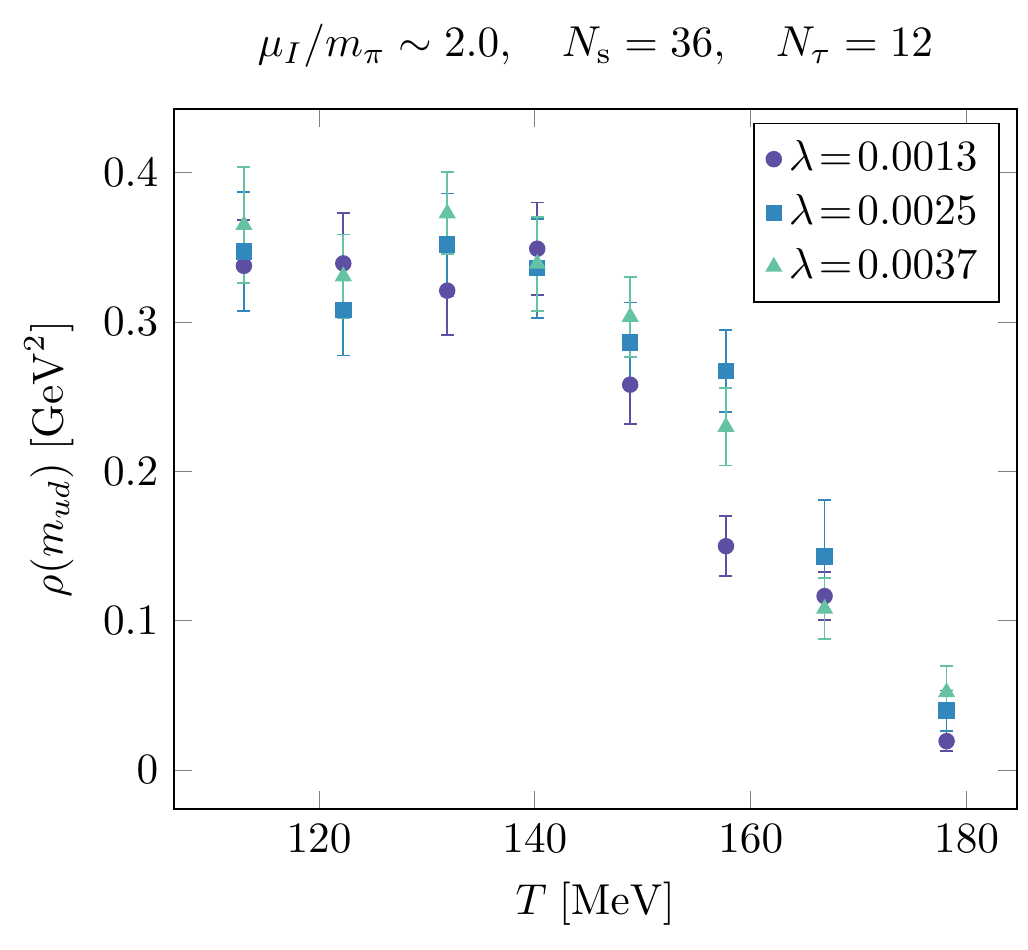}}
	\caption{\ref{fig:dataRhoMuiTDep} $\rho(\ml)$ plotted as a function of $T$ at five different (approximate) values for $\mui$, for $N_t=12$ and the smallest considered value of the pionic source $\lambda$. \ref{fig:dataRhoMuiLambdaDep} $\rho(\ml)$ as a function of $T$ for $N_t=12$, $\mui/m_\pi\sim2.0$ and three different values for $\lambda$.}
\end{figure}

Fig.~\ref{fig:dataRhoMuiTDep} shows the sensitivity of $\rho(\ml)$ to crossing the BEC boundary by raising the temperature at (approximately) fixed values of the isospin chemical potential.
It can also be seen how the interpolated spectral density is rather weakly if at all dependent on the temperature for low enough temperatures.

In Fig.~\ref{fig:dataRhoMuiLambdaDep} one can see how the obtained results for $\rho(\ml)$ are rather weakly if at all dependent on the value of the pionic source.

As compared to the preliminary results on the complex Dirac spectrum at nonzero isospin density discussed in Ref.~\cite{Brandt:2019hel}, larger volumes, finer lattices and multiple values of the pionic source have been considered for the present study.
In all considered setup, the signal shows the same qualitative trend.

\section{Discussion and conclusions}
On the basis of the presented results, we can conclude that the interpolated spectral density is undoubtedly sensitive to the BEC boundary.
No definite conclusions can, instead, be drawn yet on its sensitivity to the BEC-BCS crossover.
There a more systematic analysis is ongoing aiming at disentangling growing cutoff effects at growing $\mui$ values, from any genuine nonperturbative $\mui$-dependence as well as $T$-dependence of the BCS gap.

There are results in the literature~\cite{Son:2000xc,Cohen:2015soa} about the $\mui$-dependence of the gap as obtained using perturbation theory. $\Delta$ is found to be a rising function of $\mui$ for $\mui\geq1$ GeV, but, interestingly enough, from values that - at our energy scale - would be just a factor $~4$ smaller than the smallest values we measured for the density.
Obviously, we are rather far from that regime, but we plan simulations at larger $\mui$ values to check if the interpolated spectral density eventually starts growing with $\mui$.

It also must be kept in mind that the Banks-Casher-like relation that we have been using here as a prescription to connect the spectral density with the BCS gap is strictly valid, in the $|\mui|\gg \Lambda_{QCD}$ limit, only for $T=0$.\\

\noindent\textbf{Acknowledgements:} This work has been supported by the Deutsche Forschungsgemeinschaft (DFG, German Research Foundation) via the Emmy Noether Programme EN 1064/2-1 and TRR 211 – project number 315477589. FC also acknowledges the support by the State of Hesse within the Research Cluster ELEMENTS (Project ID 500/10.006).

\bibliographystyle{unsrt}

\begin{thebibliography}{10}

\bibitem{ValoisLat2021}
B.~B. Brandt, F.~Cuteri, G.~Endr\H{o}di, G.~Mark\'o, and A.~D.~M. Valois.
\newblock {Lattice QCD with an inhomogeneous magnetic field background}.
\newblock {\em PoS}, LATTICE2021:083, 2021.

\bibitem{Wygas:2018otj}
Mandy~M. Wygas, Isabel~M. Oldengott, Dietrich B\"odeker, and Dominik~J.
  Schwarz.
\newblock {Cosmic QCD Epoch at Nonvanishing Lepton Asymmetry}.
\newblock {\em Phys. Rev. Lett.}, 121(20):201302, 2018.

\bibitem{Middeldorf-Wygas:2020glx}
Mandy~M. Middeldorf-Wygas, Isabel~M. Oldengott, Dietrich B\"odeker, and
  Dominik~J. Schwarz.
\newblock {The cosmic QCD transition for large lepton flavour asymmetries}.
\newblock 8 2020.

\bibitem{Vovchenko:2020crk}
Volodymyr Vovchenko, Bastian~B. Brandt, Francesca Cuteri, Gergely Endr\H{o}di,
  Fazlollah Hajkarim, and J\"urgen Schaffner-Bielich.
\newblock {Pion Condensation in the Early Universe at Nonvanishing Lepton
  Flavor Asymmetry and Its Gravitational Wave Signatures}.
\newblock {\em Phys. Rev. Lett.}, 126(1):012701, 2021.

\bibitem{Son:2000xc}
D.~T. Son and Misha~A. Stephanov.
\newblock {QCD at finite isospin density}.
\newblock {\em Phys. Rev. Lett.}, 86:592--595, 2001.

\bibitem{Adhikari:2018cea}
Prabal Adhikari, Jens~O. Andersen, and Patrick Kneschke.
\newblock {Pion condensation and phase diagram in the Polyakov-loop quark-meson
  model}.
\newblock {\em Phys. Rev. D}, 98(7):074016, 2018.

\bibitem{Brandt:2017oyy}
B.~B. Brandt, G.~Endr\H{o}di, and S.~Schmalzbauer.
\newblock {QCD phase diagram for nonzero isospin-asymmetry}.
\newblock {\em Phys. Rev. D}, 97(5):054514, 2018.

\bibitem{Brandt:2018omg}
Bastian~B. Brandt and Gergely Endr\H{o}di.
\newblock {Reliability of Taylor expansions in QCD}.
\newblock {\em Phys. Rev. D}, 99(1):014518, 2019.

\bibitem{Boz:2019enj}
Tamer Boz, Pietro Giudice, Simon Hands, and Jon-Ivar Skullerud.
\newblock {Dense two-color QCD towards continuum and chiral limits}.
\newblock {\em Phys. Rev. D}, 101(7):074506, 2020.

\bibitem{Carignano:2016lxe}
Stefano Carignano, Luca Lepori, Andrea Mammarella, Massimo Mannarelli, and
  Giulia Pagliaroli.
\newblock {Scrutinizing the pion condensed phase}.
\newblock {\em Eur. Phys. J. A}, 53(2):35, 2017.

\bibitem{Kanazawa:2013crb}
Takuya Kanazawa, Tilo Wettig, and Naoki Yamamoto.
\newblock {Banks-Casher-type relation for the BCS gap at high density}.
\newblock {\em Eur. Phys. J. A}, 49:88, 2013.

\bibitem{Kogut:2002tm}
J.~B. Kogut and D.~K. Sinclair.
\newblock {Quenched lattice QCD at finite isospin density and related
  theories}.
\newblock {\em Phys. Rev. D}, 66:014508, 2002.

\bibitem{Kogut:2002zg}
J.~B. Kogut and D.~K. Sinclair.
\newblock {Lattice QCD at finite isospin density at zero and finite
  temperature}.
\newblock {\em Phys. Rev. D}, 66:034505, 2002.

\bibitem{Endrodi:2014lja}
G.~Endr\H{o}di.
\newblock {Magnetic structure of isospin-asymmetric QCD matter in neutron
  stars}.
\newblock {\em Phys. Rev. D}, 90(9):094501, 2014.

\bibitem{Borsanyi:2010cj}
Szabolcs Bors\'anyi, Gergely Endr\H{o}di, Zoltan Fodor, Antal Jakov\'ac,
  Sandor~D. Katz, Stefan Krieg, Claudia Ratti, and Kalman~K. Szab\'o.
\newblock {The QCD equation of state with dynamical quarks}.
\newblock {\em JHEP}, 11:077, 2010.

\bibitem{slepc}
Vicente Hernandez, Jose~E. Roman, and Vicente Vidal.
\newblock Slepc: A scalable and flexible toolkit for the solution of eigenvalue
  problems.
\newblock {\em ACM Trans. Math. Softw.}, 31(3):351–362, sep 2005.

\bibitem{scikit}
Fabian Pedregosa, Ga\"{e}l Varoquaux, Alexandre Gramfort, Vincent Michel,
  Bertrand Thirion, Olivier Grisel, Mathieu Blondel, Peter Prettenhofer, Ron
  Weiss, Vincent Dubourg, Jake Vanderplas, Alexandre Passos, David Cournapeau,
  Matthieu Brucher, Matthieu Perrot, and \'{E}douard Duchesnay.
\newblock Scikit-learn: Machine learning in python.
\newblock {\em J. Mach. Learn. Res.}, 12(null):2825–2830, nov 2011.

\bibitem{Brandt:2019hel}
B.~B. Brandt, F.~Cuteri, G.~Endr\H{o}di, and S.~Schmalzbauer.
\newblock {The Dirac spectrum and the BEC-BCS crossover in QCD at nonzero
  isospin asymmetry}.
\newblock {\em Particles}, 3(1):80--86, 2020.

\bibitem{Cohen:2015soa}
Thomas~D. Cohen and Srimoyee Sen.
\newblock {Deconfinement Transition at High Isospin Chemical Potential and Low
  Temperature}.
\newblock {\em Nucl. Phys. A}, 942:39--53, 2015.

\end{thebibliography}

\end{document}